\documentclass[12pt]{article}
\usepackage{epsfig,graphicx}
\usepackage{fpcp04}

\def\beq{\begin{equation}}
\def\eeq#1{\label{#1}\end{equation}}
\def\eeqn{\end{equation}}

\def\beqa{\begin{eqnarray}}
\def\eeqa#1{\label{#1}\end{eqnarray}}
\def\eeqan{\end{eqnarray}}

\let\bar=\overbar

\def\Dslash{\not{\hbox{\kern-4pt $D$}}}
\def\dslash{\not{\hbox{\kern-2pt $\del$}}}

\def\msb{{\bar{\ssstyle M \kern -1pt S}}}

\def\BB0bar{B^0 {\overline B}^0}
\def\BB0dbar{B_d^0 {\overline B}_d^0}
\def\BB0sbar{B_s^0 {\overline B}_s^0}

\RequirePackage{xspace}

\usepackage{relsize}
\def\babar{\mbox{\slshape B\kern-0.1em{\smaller A}\kern-0.1em
    B\kern-0.1em{\smaller A\kern-0.2em R}}}

\def\piz   {\ensuremath{\pi^0}\xspace}

\def\pim   {\ensuremath{\pi^-}\xspace}

\def\Kbar  {\kern 0.2em\overline{\kern -0.2em K}{}\xspace}

\def\Kz    {\ensuremath{K^0}\xspace}
\def\Kzb   {\ensuremath{\Kbar^0}\xspace}
\def\KzKzb {\ensuremath{\Kz \kern -0.16em \Kzb}\xspace}
\def\Kp    {\ensuremath{K^+}\xspace}
\def\Km    {\ensuremath{K^-}\xspace}

\def\KpKm  {\ensuremath{\Kp \kern -0.16em \Km}\xspace}
\def\KS    {\ensuremath{K^0_{\scriptscriptstyle S}}\xspace}

\def\Dbar    {\kern 0.2em\overline{\kern -0.2em D}{}\xspace}

\def\Dz      {\ensuremath{D^0}\xspace}
\def\Dzb     {\ensuremath{\Dbar^0}\xspace}
\def\DzDzb   {\ensuremath{\Dz {\kern -0.16em \Dzb}}\xspace}
\def\Dp      {\ensuremath{D^+}\xspace}
\def\Dm      {\ensuremath{D^-}\xspace}

\def\Dmp     {\ensuremath{D^\mp}\xspace}
\def\DpDm    {\ensuremath{\Dp {\kern -0.16em \Dm}}\xspace}

\def\Dstarz  {\ensuremath{D^{*0}}\xspace}

\def\Dstarp  {\ensuremath{D^{*+}}\xspace}
\def\Dstarm  {\ensuremath{D^{*-}}\xspace}
\def\Dstarpm {\ensuremath{D^{*\pm}}\xspace}

\def\Bbar    {\kern 0.18em\overline{\kern -0.18em B}{}\xspace}

\def\BB      {\ensuremath{B\Bbar}\xspace} 
\def\Bz      {\ensuremath{B^0}\xspace}
\def\Bzb     {\ensuremath{\Bbar^0}\xspace}
\def\BzBzb   {\ensuremath{\Bz {\kern -0.16em \Bzb}}\xspace}
\def\Bu      {\ensuremath{B^+}\xspace}
\def\Bub     {\ensuremath{B^-}\xspace}
\def\Bp      {\ensuremath{\Bu}\xspace}
\def\Bm      {\ensuremath{\Bub}\xspace}

\def\BpBm    {\ensuremath{\Bu {\kern -0.16em \Bub}}\xspace}

\mathchardef\Upsilon="7107
\def\Y#1S{\ensuremath{\Upsilon{(#1S)}}\xspace}

\mathchardef\Deltares="7101
\mathchardef\Xi="7104
\mathchardef\Lambda="7103
\mathchardef\Sigma="7106
\mathchardef\Omega="710A

\def\Deltabar{\kern 0.25em\overline{\kern -0.25em \Deltares}{}\xspace}
\def\Lbar{\kern 0.2em\overline{\kern -0.2em\Lambda\kern 0.05em}\kern-0.05em{}\xspace}
\def\Sigbar{\kern 0.2em\overline{\kern -0.2em \Sigma}{}\xspace}
\def\Xibar{\kern 0.2em\overline{\kern -0.2em \Xi}{}\xspace}
\def\Obar{\kern 0.2em\overline{\kern -0.2em \Omega}{}\xspace}
\def\Nbar{\kern 0.2em\overline{\kern -0.2em N}{}\xspace}
\def\Xb{\kern 0.2em\overline{\kern -0.2em X}{}\xspace}

\newcommand{\tev}{\ensuremath{\mathrm{\,Te\kern -0.1em V}}\xspace}
\newcommand{\gev}{\ensuremath{\mathrm{\,Ge\kern -0.1em V}}\xspace}
\newcommand{\mev}{\ensuremath{\mathrm{\,Me\kern -0.1em V}}\xspace}
\newcommand{\kev}{\ensuremath{\mathrm{\,ke\kern -0.1em V}}\xspace}
\newcommand{\ev}{\ensuremath{\mathrm{\,e\kern -0.1em V}}\xspace}
\newcommand{\gevc}{\ensuremath{{\mathrm{\,Ge\kern -0.1em V\!/}c}}\xspace}
\newcommand{\mevc}{\ensuremath{{\mathrm{\,Me\kern -0.1em V\!/}c}}\xspace}
\newcommand{\gevcc}{\ensuremath{{\mathrm{\,Ge\kern -0.1em V\!/}c^2}}\xspace}
\newcommand{\mevcc}{\ensuremath{{\mathrm{\,Me\kern -0.1em V\!/}c^2}}\xspace}

\def\invfb   {\ensuremath{\mbox{\,fb}^{-1}}\xspace}

\def\mus  {\ensuremath{\rm \,\mus}\xspace}

\def\mus        {\ensuremath{\,\mu{\rm s}}\xspace}    

\def\to                 {\ensuremath{\rightarrow}\xspace}

\def\pep2{PEP-II}

\def\gsim{{~\raise.15em\hbox{$>$}\kern-.85em
          \lower.35em\hbox{$\sim$}~}\xspace}
\def\lsim{{~\raise.15em\hbox{$<$}\kern-.85em
          \lower.35em\hbox{$\sim$}~}\xspace}

\def\CP                {\ensuremath{C\!P}\xspace}

\xspace

\def\jetset74   {\mbox{\tt Jetset \hspace{-0.5em}7.\hspace{-0.2em}4}\xspace}

\begin{document}

\Title{\boldmath Measurement of $\gamma$ and $2\beta + \gamma$}
\bigskip

%
\label{JAlbertStart}

%
\author{ Justin Albert\index{Albert, J.} }

%
\address{
(from the \babar\ Collaboration)\\
Department of Physics\\
California Institute of Technology\\
Pasadena, CA 91125, USA\\
}

\makeauthor\abstracts{
We report on the initial measurements of the angle $\gamma$ and the sum of angles 
$2\beta + \gamma$ of the Unitarity Triangle.  When compared with indirect
information on the value of $\gamma$ from other measurements of CKM parameters,
the measurement of these angles
will provide a precise test of Standard Model predictions, as statistics increase.
There are several methods for directly measuring $\gamma$ and $2\beta + \gamma$.
We report on the status of each of these techniques, and the resulting constraints
on the values of these angles.
}

\section{Introduction}
The comparison of measurements of the angles and sides of the Unitarity Triangle
provides a test of the Standard Model, in which \CP-violation is solely due to a single
complex phase in $V$, the Cabbibo-Kobayashi-Maskawa (CKM) quark mixing matrix.
The angle $\gamma \equiv \arg[-V_{\rm cd}V^{*}_{\rm cb}/V_{\rm ud}V^{*}_{\rm ub}]$ 
is considered to be the most difficult to measure of the three Unitarity Triangle angles.
The difficulty is due to the fact that the interference terms that provide the sensitivity
to $\gamma$ tend to be small, due to small branching fractions, lower
reconstruction efficiencies than with typical charmonium or charmless $B$ decays, and 
relevant magnitudes of interfering amplitudes that are far from equal.

Nevertheless, there exist several techniques for directly measuring $\gamma$ and 
$2\beta + \gamma$.  These techniques can be divided into three classes: those that
use a time-independent \CP asymmetry between color-allowed $B \to \Dz K$ and 
color-suppressed $B \to \Dzb K$ amplitudes to directly measure $\gamma$, which is
the relative weak phase between these amplitudes; those that use a time-dependent
asymmetry between favored and suppressed $B \to D\pi$ or $B \to \Dz K^0$ amplitudes;
and a third type of technique, which uses a combination of
time-dependent and time-independent asymmetries and branching fractions
in $B \to D_{(s)}^{(*)}D^{(*)}$ to solve for the value of $\gamma$.

\section{Time-Independent Techniques}

The time-independent techniques each use an interference between color-allowed $B \to \Dz K$ and
color-suppressed $B \to \Dzb K$ amplitudes to constrain $\gamma$ through a $CP$-violating asymmetry in
time-integrated decay rates.  The $\Dz K$ and $\Dzb K$ amplitudes have a relative weak phase of $\gamma$, but one always needs
two more pieces of experimental information to form a constraint: the relative magnitude of the amplitudes
$r_B \equiv |\frac{A(b \to u)}{A(b \to c)}|$ and the strong phase difference between the two amplitudes
$\delta_B$.  Naturally, the larger the interference, \textit{i.e.} the closer $r_B$ is to 1, the better
the constraint will be on the value of $\gamma$ for a given dataset.  
\begin{figure}[tb]
\begin{center}
\epsfig{file=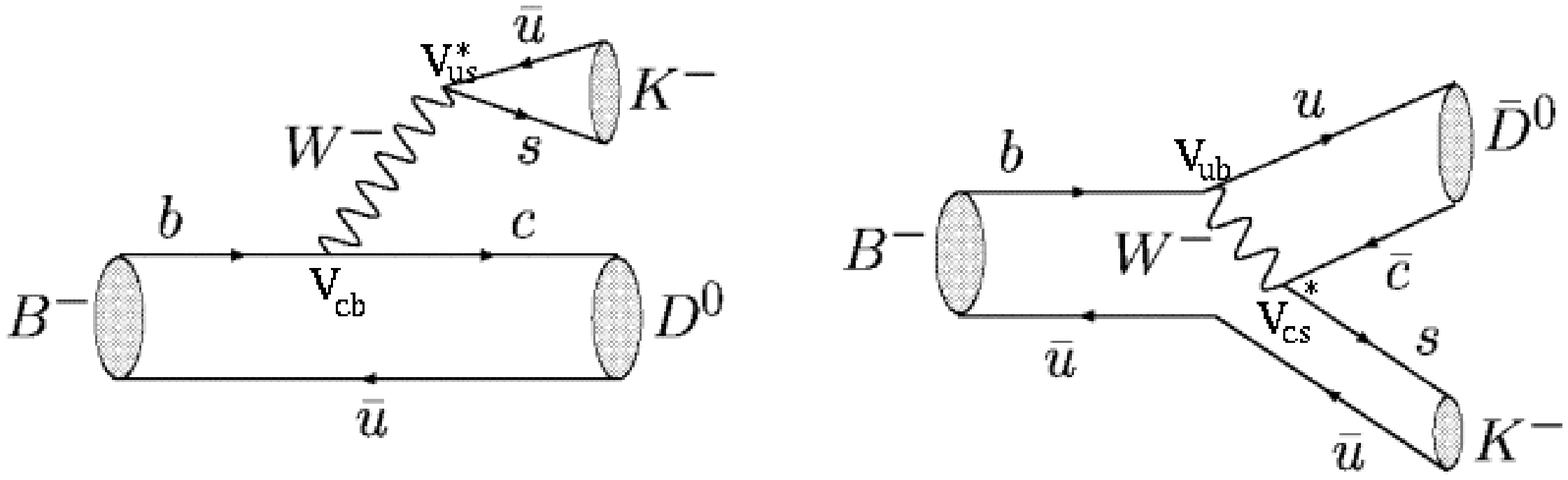,height=5.2cm}
\caption{The $B \to \Dz K$ and $B \to \Dzb K$ amplitudes have a relative weak phase of 
$\gamma$.}
\label{fig:DKinterfere}
\end{center}
\end{figure}

The original technique for measuring $\gamma$, the Gronau-London-Wyler
(GLW) method, uses an asymmetry between 
$\Bm \to \Dz \Km$ and $\Bp \to \Dzb \Kp$, where the \Dz (\Dzb) decays to a \CP
eigenstate mode ($\pi^{+}\pi^{-}$, $K^{+}K^{-}$, $\KS\piz$, $\KS\phi$, or $\KS\omega$)~\cite{GLW}.  There are 4 observables:
\begin{eqnarray}
R_{\CP^{\pm}} & \equiv & \frac{\Gamma(\Bm \to D^{0}_{\CP^{\pm}}K^-) + \Gamma(\Bp \to D^{0}_{\CP^{\pm}}K^+)}{2\Gamma(\Bm \to D^{0}_{\rm flav}K^-)} =
                         1 \pm 2r_{B}\cos\gamma\cos\delta_{B} + r_{B}^2\nonumber\\
A_{\CP^{\pm}} & \equiv & \frac{\Gamma(\Bm \to D^{0}_{\CP^{\pm}}K^-) - \Gamma(\Bp \to D^{0}_{\CP^{\pm}}K^+)}{\Gamma(\Bm \to D^{0}_{\CP^{\pm}}K^-) + \Gamma(\Bp \to D^{0}_{\CP^{\pm}}K^+)} =
                         \pm 2r_{B}\sin\gamma\sin\delta_{B}/R_{\CP^{\pm}}\nonumber
\end{eqnarray}
(where $\CP^{+}$ refers to the \CP-even final states $\pi^{+}\pi^{-}$ and $K^{+}K^{-}$ and $\CP^{-}$ refers 
to the \CP-odd final states $\KS\piz$, $\KS\phi$, and $\KS\omega$)
thus allowing a solution of the 3 unknowns ($r_B$, $\delta_B$, and $\gamma$), up to an
8-fold ambiguity in $\gamma$ (when no external prior is taken for the value of $\delta_B$).
In the GLW technique, as well as in the two other time-independent techniques,
combining information from both $B \to \Dz K$ and $B \to \Dstarz K$ final states, and reconstructing both
the $\Dstarz \to \Dz\piz$ and the $\Dstarz \to \Dz\gamma$ final states, allows further constraint beyond
that from a single final state, partially due to a $180^\circ$ phase difference between the $\Dz\piz$ and $\Dz\gamma$ final states~\cite{SoniBondar}.

The GLW method is theoretically clean, with nearly no hadronic uncertainty.  However it is experimentally challenging,
as the branching fractions to the relevant final states are small: the $B \to DK$ branching fractions are at the 
$10^{-4}$ level, the branching fractions of the $D$ into \CP eigenstate modes are of order $10^{-2}$, and the overall detection
efficiencies are around 25\%.  Thus, using a sample of $214 \times 10^{6}$ \BB events, \babar's yield for the GLW final states 
is $93 \pm 15$ for $B \to D^{0}K$ where the \Dz 
decays to the two \CP-even final states and $76 \pm 13$ for \Dz decays to the \CP-odd
final state.  \babar\ also reconstructs $B \to D^{0}K^{*}$ (with $K^{*} \to \KS\pim$), and obtains yields of 
$34.4 \pm 6.9$ events in the \CP-even modes and $15.1 \pm 5.8$ events in the \CP-odd modes of the \Dz in a sample
of $227 \times 10^{6}$ \BB events~\cite{babarGLWDK,babarGLWDstK}.
\begin{figure}[tb]
\begin{center}
\epsfig{file=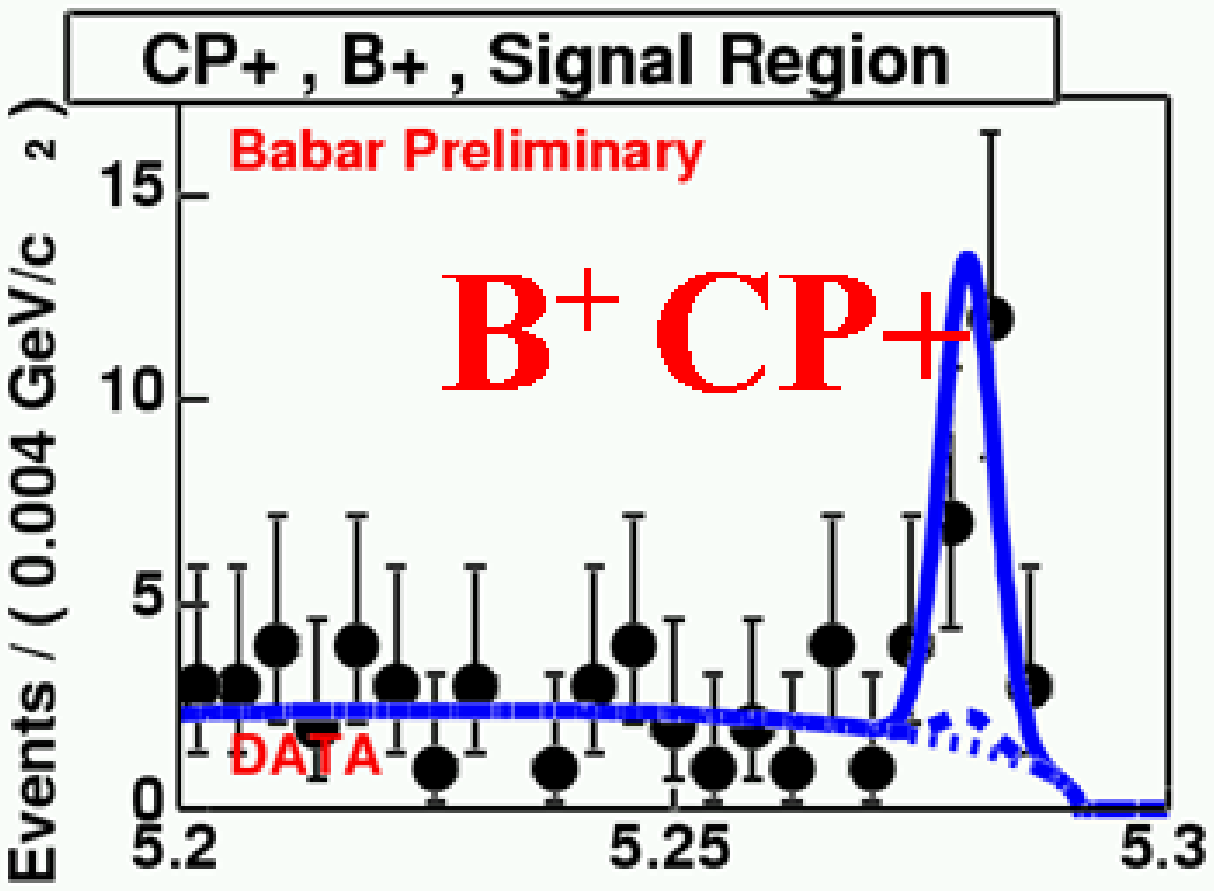,height=3.0cm}
\epsfig{file=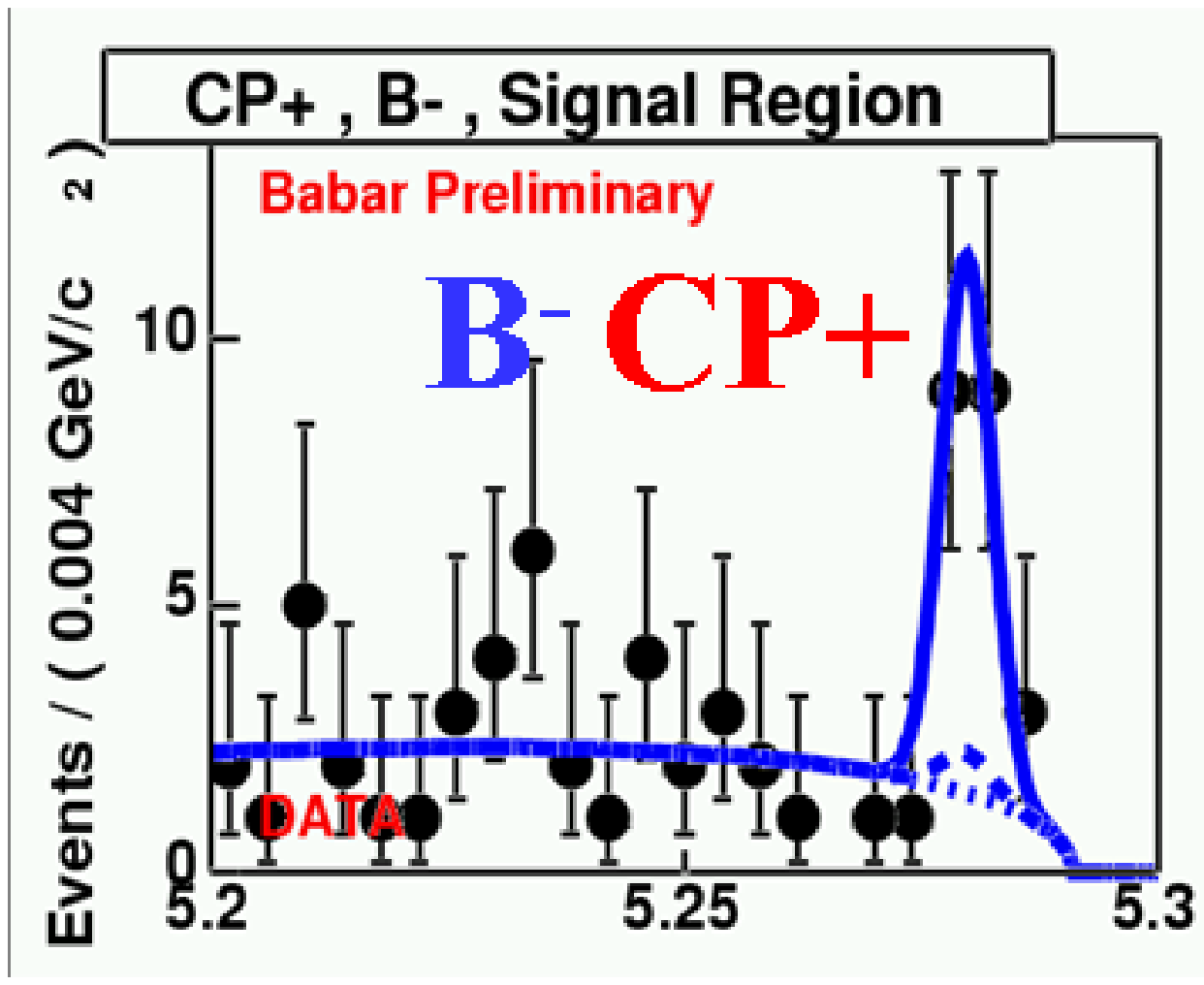,height=3.0cm}
\epsfig{file=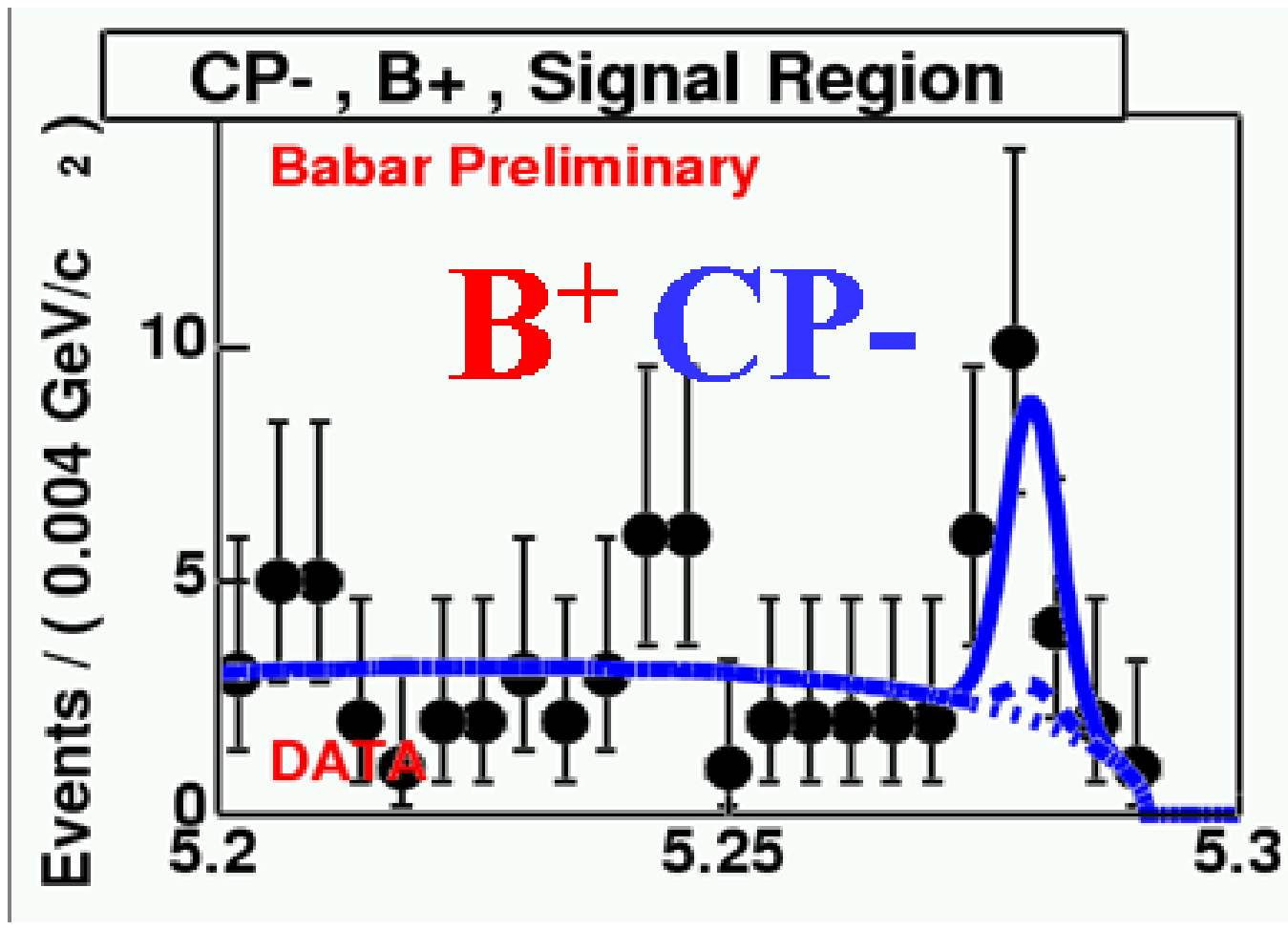,height=3.0cm}
\epsfig{file=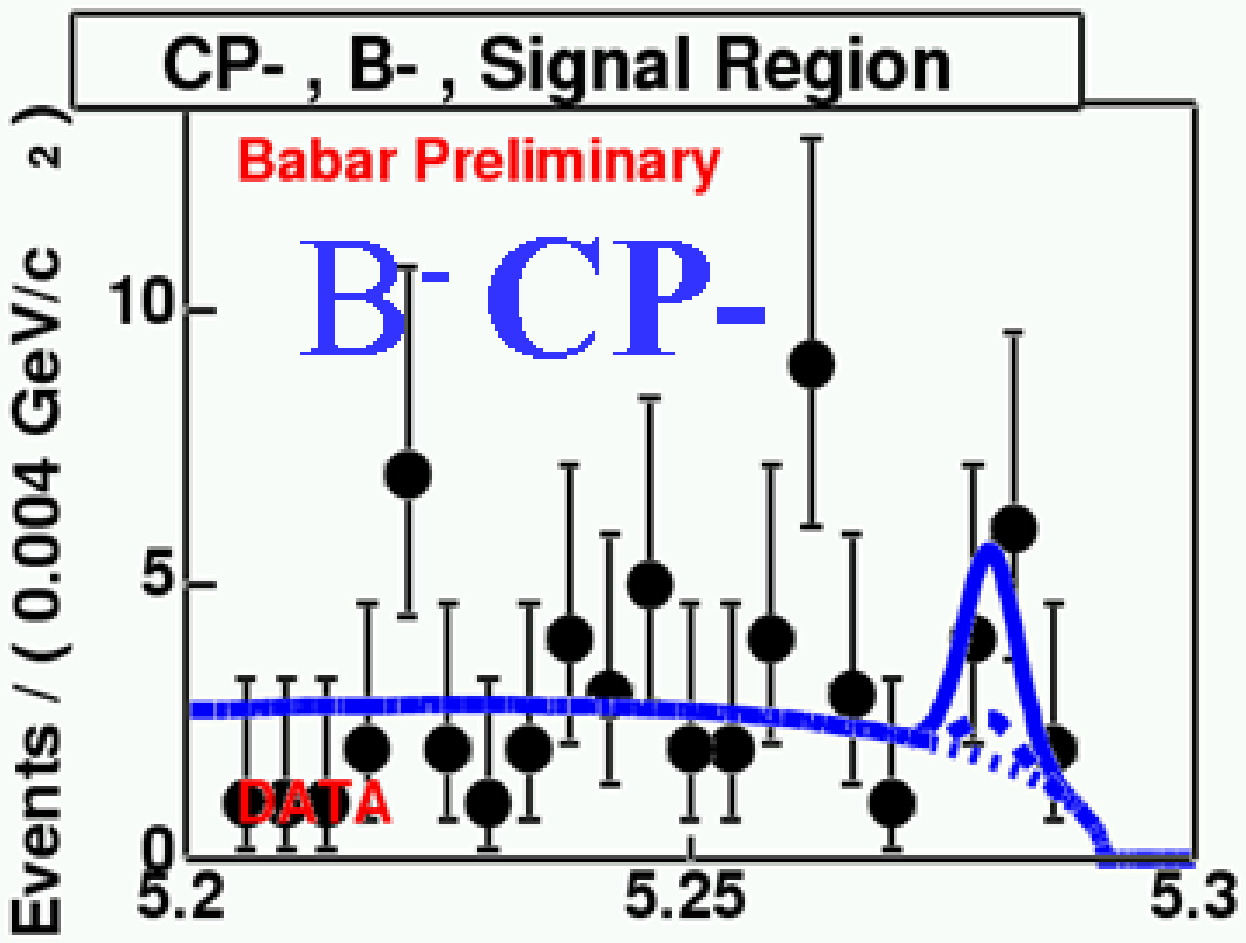,height=3.0cm}
\caption{Yields at \babar\ in a sample of $227 \times 10^{6}$ \BB events
for $B \to D^{0}K^{*}$ (with $K^{*} \to \KS\pim$).  The left two plots show \Bp and \Bm
yields to the \CP-even \Dz decay modes $\pi^{+}\pi^{-}$ and $K^{+}K^{-}$ (a total of $34.4 \pm 6.9$ events)
and the right two show the yield to the \CP-odd 
modes $\KS\piz$, $\KS\phi$, and $\KS\omega$ ($15.1 \pm 5.8$ events)~\cite{babarGLWDstK}.}
\label{fig:GLW_DKst}
\end{center}
\end{figure}

Using the above event yields, the 4 GLW experimental observables are measured at \babar\ to be 
$R_{\CP^{+}} = 0.87 \pm 0.14 \pm 0.06$,
$A_{\CP^{+}} = 0.40 \pm 0.15 \pm 0.08$,
$R_{\CP^{-}} = 0.80 \pm 0.14 \pm 0.08$,
$A_{\CP^{-}} = 0.21 \pm 0.17 \pm 0.07$ for the $B \to D^{0}K$ modes, and
$R_{\CP^{+}} = 1.77 \pm 0.37 \pm 0.12$,
$A_{\CP^{+}} = -0.09 \pm 0.20 \pm 0.06$,
$R_{\CP^{-}} = 0.76 \pm 0.29 \pm 0.06^{-0.04}_{-0.14}$,
$A_{\CP^{-}} = -0.33 \pm 0.34 \pm 0.10 (+0.15 \pm 0.10) \cdot (A_{\CP^{-}} - A_{\CP^{+}})$ for the $B \to D^{0}K^{*}$ modes,
where the third uncertainty in the last two measurements reflects possible interference effects in final states with $\phi$ and $\omega$~\cite{babarGLWDK,babarGLWDstK}.
\babar\ also measures $R_{\CP^{+}} = 0.88 \pm 0.26^{+0.10}_{-0.08}$ and
$A_{\CP^{+}} = -0.02 \pm 0.24 \pm 0.05$ in $B \to D^{*0}K$ events, with $D^{*0} \to D^{0}\pi^0$, 
in a sample of 
$123 \times 10^{6}$ \BB events~\cite{babarGLWold}.

Belle measures 
$R_{\CP^{+}} = 0.98 \pm 0.18 \pm 0.10$,
$A_{\CP^{+}} = 0.07 \pm 0.14 \pm 0.06$,
$R_{\CP^{-}} = 1.29 \pm 0.16 \pm 0.08$,
$A_{\CP^{-}} = -0.11 \pm 0.14 \pm 0.05$ for the $B \to D^{0}K$ modes, and
$R_{\CP^{+}} = 1.43 \pm 0.28 \pm 0.06$,
$A_{\CP^{+}} = -0.27 \pm 0.25 \pm 0.04$,
$R_{\CP^{-}} = 0.94 \pm 0.28 \pm 0.06$,
$A_{\CP^{-}} = 0.26 \pm 0.26 \pm 0.03$ for $B \to D^{*0}K$ ($D^{*0} \to D^{0}\pi^0$) modes, each
in a sample of 250 \invfb of data~\cite{belleGLW}.

Using the $B \to D^{0}K^{*}$ results, \babar\ constrains the value of the theoretical parameter 
$r_B^2$ to be $0.23 \pm 0.24$.  However, more statistics are needed to constrain $\gamma$ from this method.

The Atwood-Dunietz-Soni (ADS) method also uses interference between color-allowed $B \to \Dz K$ and
color-suppressed $B \to \Dzb K$ amplitudes, but instead of using \CP eigenstate decays of the \Dz,
the decays \Dz $\to K^{+}\pi^-$ and \Dzb $\to K^{+}\pi^-$ are used~\cite{ADS}.  There are two 
observables:
\begin{eqnarray}
R_{\rm ADS} & \equiv & \frac{\Gamma(\Bm \to D^{0}(\to K^{+}\pi^{-})K^-) + \Gamma(\Bp \to D^{0}(\to K^{-}\pi^{+})K^+)}{\Gamma(\Bm \to D^{0}(\to K^{-}\pi^{+})K^-) + \Gamma(\Bp \to D^{0}(\to K^{+}\pi^{-})K^+)}\nonumber\\
            &    =   &       r_{D}^2 + 2r_{B}r_{D}\cos\gamma\cos(\delta_{B} + \delta_{D}) + r_{B}^2\nonumber\\
A_{\rm ADS} & \equiv & \frac{\Gamma(\Bm \to D^{0}(\to K^{+}\pi^{-})K^-) - \Gamma(\Bp \to D^{0}(\to K^{-}\pi^{+})K^+)}{\Gamma(\Bm \to D^{0}(\to K^{+}\pi^{-})K^-) + \Gamma(\Bp \to D^{0}(\to K^{-}\pi^{+})K^+)}\nonumber\\
            &    =   &       2r_{B}r_{D}\sin\gamma\sin(\delta_{B} + \delta_{D})/R_{\rm ADS}\nonumber
\end{eqnarray}
The value of $r_{D} \equiv \frac{|A(D^{0} \to K^{+}\pi^{-})|}{|A(D^{0} \to K^{-}\pi^{+})|}$ is constrained to the experimental value of $0.060 \pm 0.003$~\cite{ADSrD}.  The
values of $r_{B}$ and $\delta_{B}$ are equal to those from the GLW analysis described above.  One is left with the two theoretical unknowns $\delta_{D}$ and $\gamma$, which
can in principle be determined from the two experimental observables.  As in the GLW and Dalitz techniques (the latter of which will be discussed following this one), the ADS
technique benefits from combination of information from $B \to \Dz K$ and $B \to \Dstarz K$~\cite{SoniBondar}.

, and reconstructing both 
the $\Dstarz \to \Dz\piz$ and the $\Dstarz \to \Dz\gamma$ final states, allows further constraint beyond
that from a single final state, partially due to a $180^\circ$ phase difference between the $\Dz\piz$ and $\Dz\gamma$ final states.  

\begin{figure}[tb]
\begin{center}
\epsfig{file=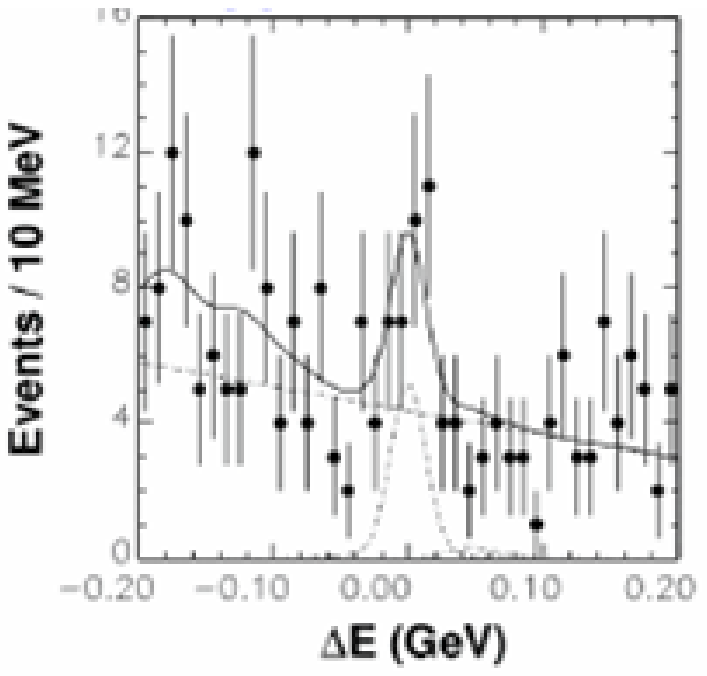,height=4.0cm}
\epsfig{file=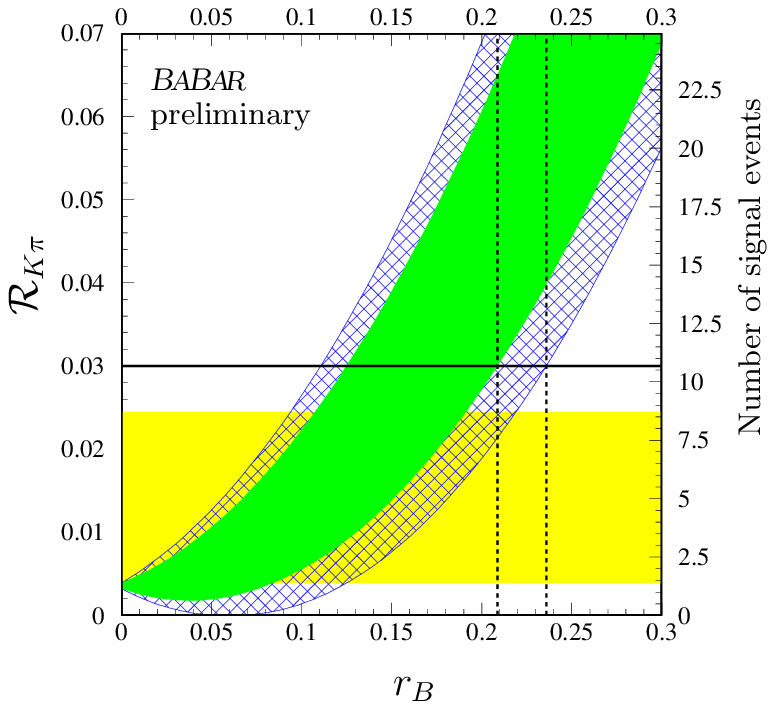,height=4.0cm}
\caption{(Left) Yield at Belle for the ADS decay mode $B \to D^{0}(\to K\pi)K$ with opposite-sign kaons.  Belle 
establishes a limit on this branching fraction of $7.6 \times 10^{-7}$ at 90\% c.l.~\cite{belleADS}  (Right) Constraints in the 
$r_B$-$R_{\rm ADS}$
plane from \babar.  The green region allows any value of $\delta_D$ and $48^\circ < \gamma < 73^\circ$, and the
hatched region allows any value of $\gamma$.  \babar\ constrains $r_B < 0.23$ at 90\% c.l.~\cite{babarADS}}
\label{fig:ADS}
\end{center}
\end{figure}

Similar to the GLW analysis, the ADS analysis is theoretically clean but suffers from highly suppressed decay rates into the relevant final states.  
Using a sample of $227 \times 10^{6}$ \BB events, \babar\ reconstructs $4.7^{+4.0}_{-3.2}$ events
in the $\Bm \to D^{0}K^{-}$ $(\Dz \to K^{+}\pi^{-})$ channel, $-0.2^{+1.3}_{-0.8}$ events in the $\Bm \to D^{*0}K^{-}$ $(D^{*0} \to D^{0}\pi^{0}, \Dz \to K^{+}\pi^{-})$ channel,
and $1.2^{+2.1}_{-1.4}$ events in the $\Bm \to D^{*0}K^{-}$ $(D^{*0} \to D^{0}\gamma, \Dz \to K^{+}\pi^{-})$ 
channel~\cite{babarADS}.  
No significant signal is seen for any of these channels.
Belle reconstructs $14.7 \pm 7.6$ events in the first of these channels, 
also not significant~\cite{belleADS}.  Using these values, \babar\ constrains $R_{\rm ADS}$ from $\Bm \to D^{0}K^{-}$
channel to be less than 0.030 at 90\% confidence level (c.l.).  From this result, and allowing any value of $\delta_D$ and $\gamma$, one can constrain $r_B$ to be less than 0.23 at
90\% c.l.~as shown in Fig.~\ref{fig:ADS} (right).  Belle similarly constrains $R_{\rm ADS} < 0.047$ and $r_B < 0.28$, both at 90\% 
c.l.  However, similar to the GLW analysis,
more statistics are needed to constrain $\gamma$ from the ADS method.

A third method to constrain $\gamma$ using interference between $B \to \Dz K$ and $B \to \Dzb K$ amplitudes is the
Dalitz technique~\cite{DalitzSoniEtc}, which uses \Dz and \Dzb decays to the common final state $K_{S}\pi^{+}\pi^{-}$.  By simultaneously fitting
the Dalitz distributions of the \Dz $\to K_{S}\pi^{+}\pi^{-}$ decays from \Bp and \Bm data, one can constrain the values of
the three theoretical unknowns ($r_B,\delta_B,\gamma$)~\cite{Dalitz}.

\begin{figure}[tb]
\begin{center}
\epsfig{file=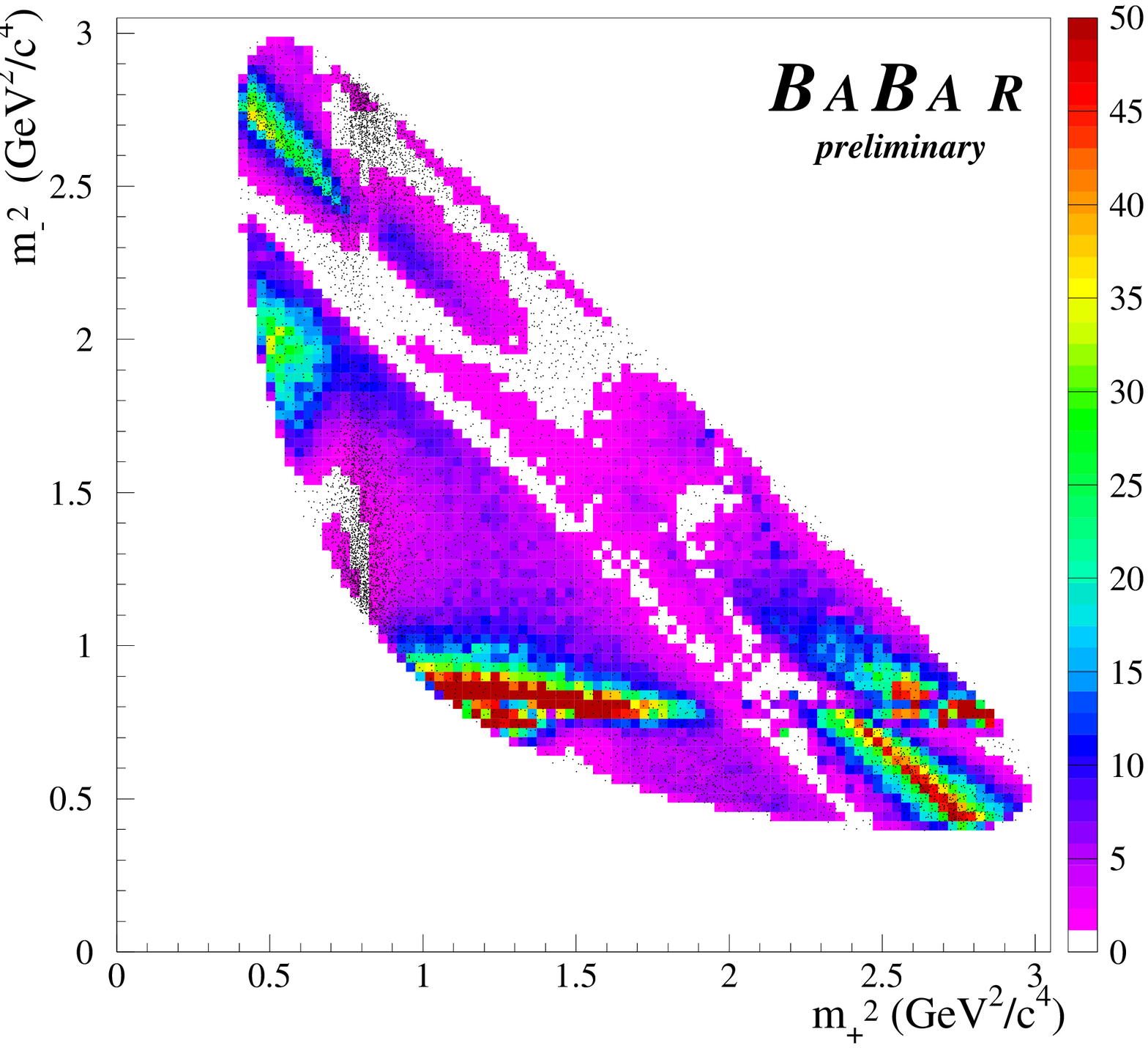,height=3.3cm}
\epsfig{file=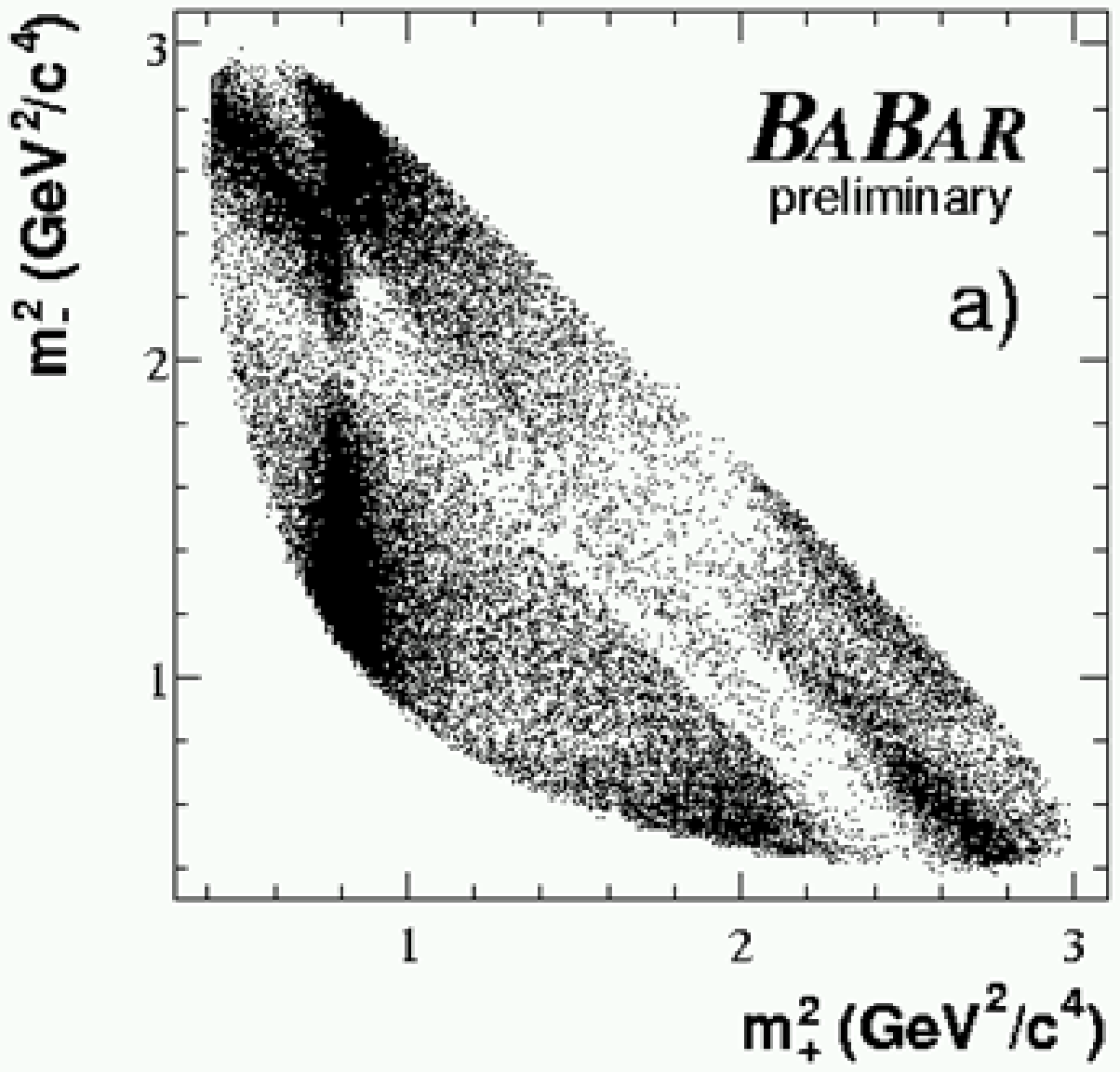,height=3.3cm}
\epsfig{file=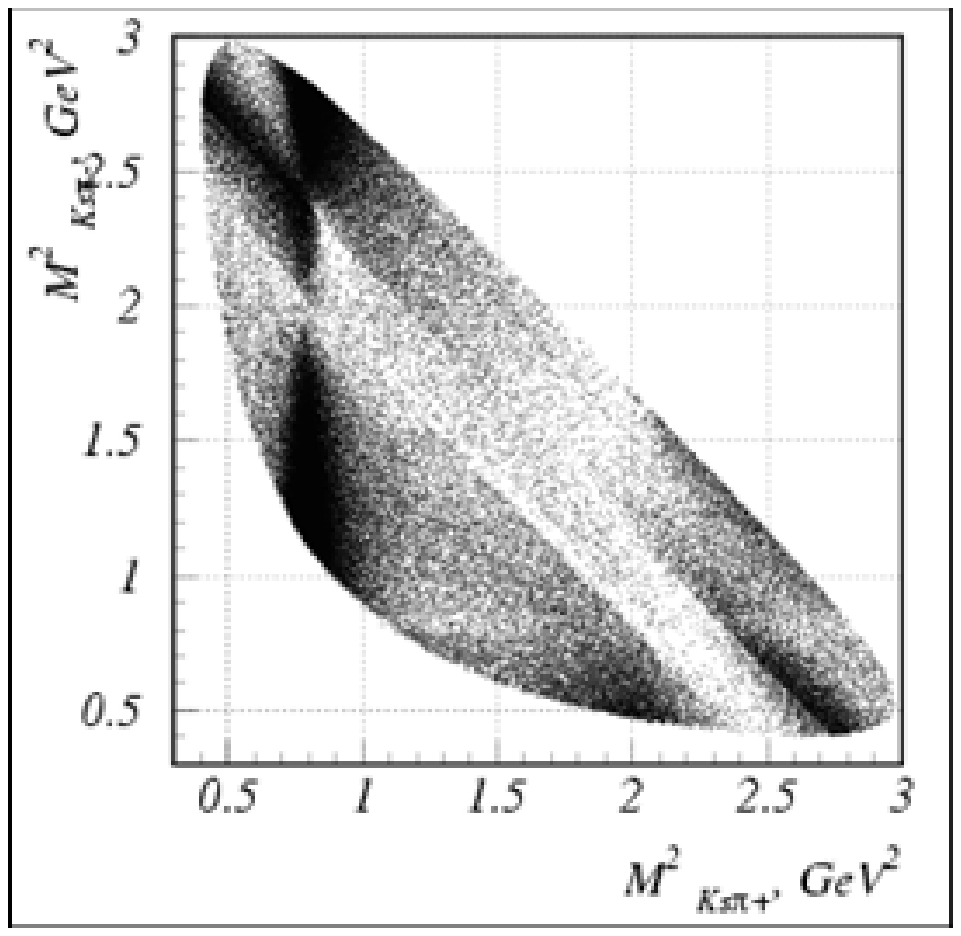,height=3.3cm}

\epsfig{file=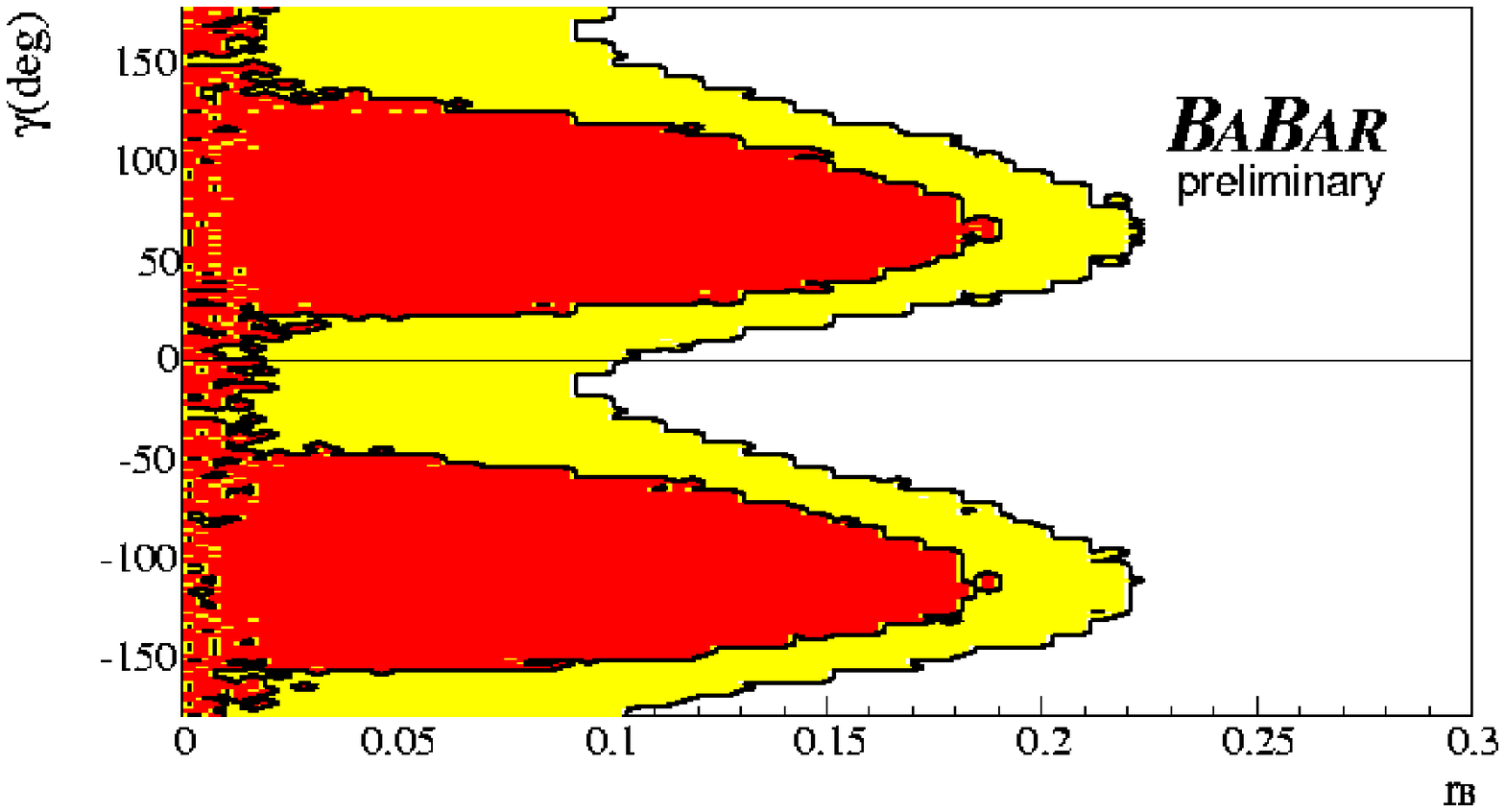,height=3.3cm}
\epsfig{file=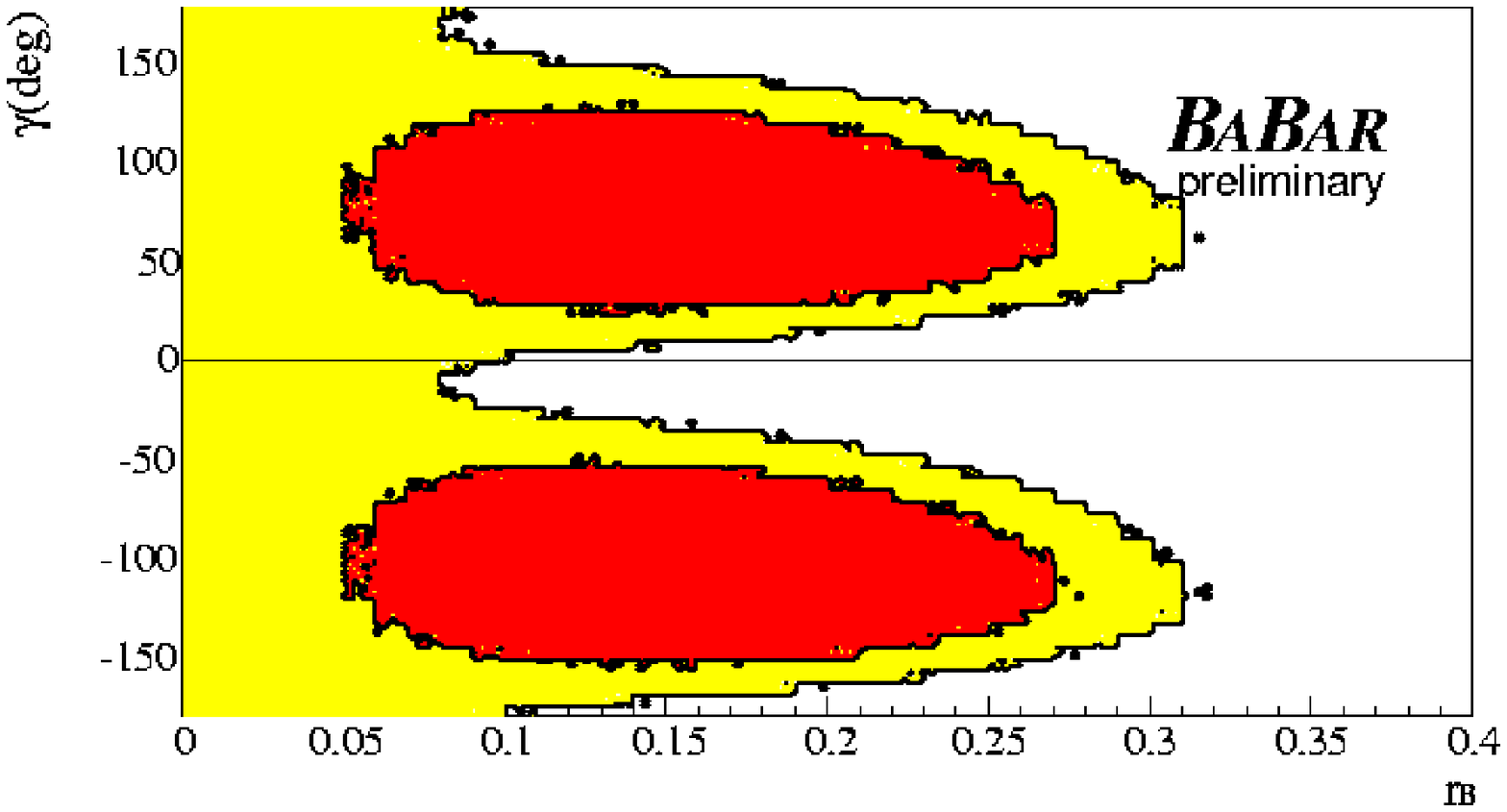,height=3.3cm}
\epsfig{file=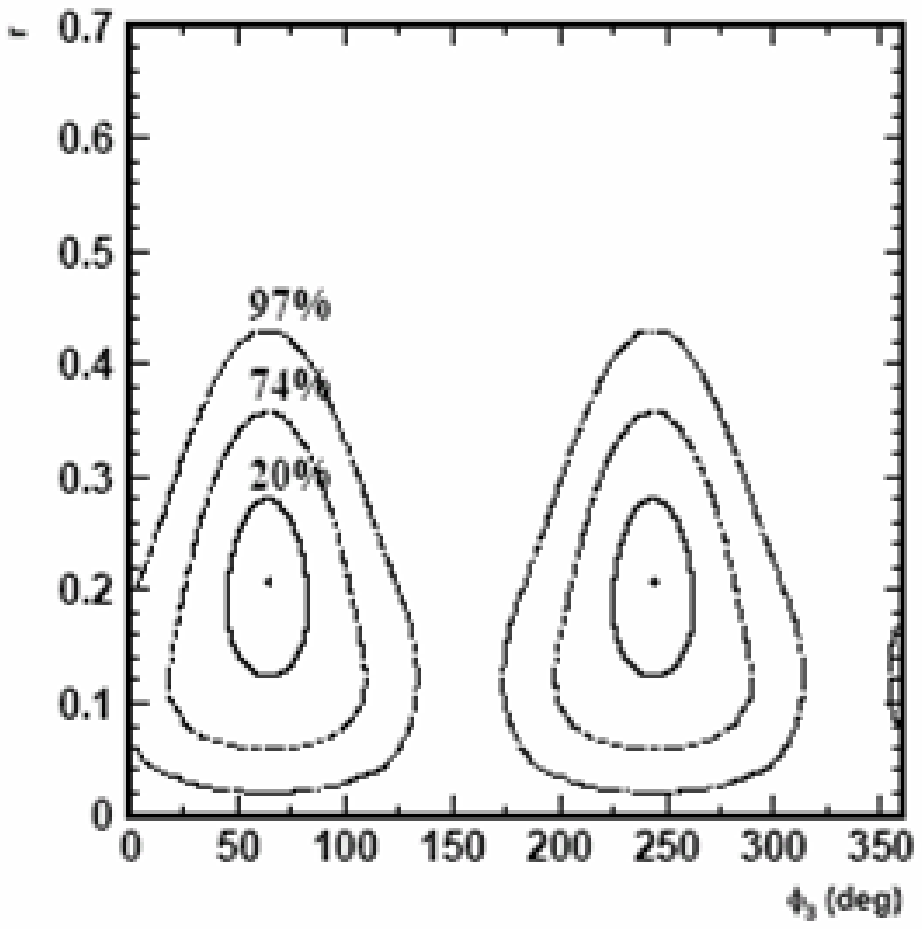,height=3.3cm}
\caption{
(Upper left) Sensitivity to $\gamma$ for $B \to \Dz K$ events with \Dz $\to K_{S}\pi^{+}\pi^{-}$ as a 
function of position within the Dalitz plot.  (Upper middle and right) Dalitz distributions obtained
using an inclusive sample of \Dz $\to K_{S}\pi^{+}\pi^{-}$ events from \babar\ and Belle 
respectively~\cite{babarDalitz,belleDalitz}.
Resulting constraints on the $r_B$-$\gamma$ plane at 68\% (red) and 90\% (yellow) confidence levels
from (lower left) $B \to \Dz K$ and (lower middle) $B \to \Dstarz K$.
(Lower right) Constraints on the $\gamma$-$r_B$ plane from Belle at 20\%, 74\%, and 90\% confidence levels.
}
\label{fig:Dalitz}
\end{center}
\end{figure}

One must first constrain the value of the strong phase $\delta_D$ and the ratio of magnitudes of interfering amplitudes $r_D$ of the \Dz decay, parameters which are a function of position 
in the Dalitz plot.  The sample of $B \to \Dz K$ decays alone does not contain nearly enough statistics to constrain these functions.  However,
a sample of inclusive \Dz $\to K_{S}\pi^{+}\pi^{-}$ decays is of order $10^{4}$ times larger, and provides sufficient statistics to constrain
the $\delta_D$ and $r_D$ variation.  
\babar\ perfoms a fit to 16 different resonances, plus a non-resonant component, to the inclusive 
\Dz $\to K_{S}\pi^{+}\pi^{-}$ Dalitz distribution, and determines the amplitudes and relative phases of each, allowing the calculation
of the $\delta_D$ and $r_D$ variation.  
Belle performs a similar Dalitz fit.  The distributions from \babar\ and Belle are shown in 
Fig.~\ref{fig:Dalitz}~\cite{babarDalitz,belleDalitz}.

Once $\delta_D$ and $r_D$ are known, one can then observe the exclusive channels to constrain $r_B$, $\delta_B$, and $\gamma$.
Using a sample of $211 \times 10^{6}$ \BB events, \babar\ reconstructs $261 \pm 19$ events in the $B \to \Dz 
(\to K_{S}\pi^{+}\pi^{-}) K$
channel.  \babar\ also reconstructs the decays 
$B \to \Dstarz (\to D^{0}x, D^{0} \to K_{S}\pi^{+}\pi^{-}) K$, with $x = \piz$ or $\gamma$, 
and obtains $83 \pm 11$ and $40 \pm 8$ events in those channels respectively.  Belle reconstructs $209 \pm 16$ and $58 \pm 8$ events
in 250 \invfb for the first two of these three channels (Belle does not presently reconstruct the $\Dstarz \to D^{0}\gamma$ mode).  

Using a simultaneous fit to the \Bp and \Bm Dalitz distributions, \babar\ measures $\delta_B = (114 \pm 41 \pm 8 \pm 10)^\circ$
and $\delta^{*}_B = (123 \pm 34 \pm 14 \pm 10)^\circ$,
where in both cases there is a $+(0^\circ, 180^\circ)$ ambiguity and
the first error is statistical, the second is systematic, and the third is from uncertainty on the phase 
variation model.
\babar\ also finds $r_B^{*} = 0.155^{+0.070}_{-0.077} \pm 0.040 \pm 0.020$ and
constrains $r_B$ to be less than 0.19 at 90\% c.l.  The value of $\gamma$ is constrained by \babar, from 
a combined fit to the the $\Dz K$ and $\Dstarz K$ modes,
to be $(70 \pm 26 \pm 10 \pm 10)^\circ$.  

As noted above, the uncertainty on the value of $\gamma$ for each of the time-independent techniques strongly depends on the value of $r_B$; a larger value of 
this parameter implies a larger $\Dz K$--$\Dzb K$ interference term, thus a smaller uncertainty on the measured value of 
$\gamma$.  
In each of the three analyses above, Belle reports a larger central value of $r_B$ than \babar.  In the case of the GLW and Dalitz
analyses, Belle's central values both are greater than the 90\% c.l.~upper limits on $r_B$ placed by \babar.  While none of the inconsistencies
are, by themselves, statistically significant, it is unclear why this trend has so far occured in each of the above analyses.

For the Dalitz analysis, Belle reports $r_B = 0.21 \pm 0.08 \pm 0.03 \pm 0.04$, $\delta_B = (64 \pm 19 \pm 13 \pm 11)^\circ$, and $\gamma = (64 \pm 19 \pm 13 \pm 11)^\circ$.
The smaller uncertainty on $\gamma$, as compared with the \babar\ analysis, is due to the apparent larger central value of $r_B$ that Belle reconstructs.
(This value has in fact declined from Belle's previous measurement of $r_B = 0.26^{+0.10}_{-0.14} \pm 0.03 \pm 0.04$, using an 
earlier sample of 140 \invfb of data~\cite{belleDalitzold}.  The 
declining value of $r_B$ appears to explain why, after increasing their data sample by over a factor of two, Belle's uncertainty on 
$\gamma$ 
has actually increased slightly.)

\section{\boldmath Measuring $\sin(2\beta + \gamma)$ Using Time-Dependent Asymmetries}

Time-dependent asymmetries in \Bz $\to D^{(*)}\pi$, $D^{(*)}\rho$, and $D^{(*)0}(\bar{D}^{(*)0})K^{*0}$ can provide information on the value of $\sin(2\beta + \gamma)$ --- and
thus the value of $\gamma$, since $\beta$ is so well-constrained.  The  \Bz $\to D^{(*)}\pi$ and $D^{(*)}\rho$ methods use an interference between the usual
Cabibbo-favored $b \to c$ channel and the doubly-Cabibbo-suppressed $b \to u$ 
channel~\cite{sin2bgdpirho}.  These two amplitudes have a relative weak phase of $\gamma$, and a weak
phase of $2\beta$ is provided by the \BzBzb mixing.  As the amplitude for the $b \to u$ channel is very small compared with $b \to c$, the 
time-dependent asymmetry is a small effect, of order $\lambda^2$.  

There are two experimental methods for reconstructing \Bz(\Bzb) $\to D^{(*)}\pi$ and $D^{(*)}\rho$ decays to determine $\gamma$.  One can perform exclusive reconstruction, where
one fully reconstructs all the final state particles for the low-multiplicity hadronic $D^{(*)}$ decay modes.  This method has a very high signal purity (typically near 90\%),
but one cannot reconstruct the majority of the $D^{(*)}$ decays, \textit{i.e.} those into semi-leptonic or high-multiplicity hadronic decay modes.  In order to obtain a higher
efficiency, one can perform partial reconstruction of $D^{*}\pi$ and $D^{*}\rho$, by reconstructing only the slow pion, and not the \Dz, from the $D^{*} \to D^{0}\pi$ decay.  Using
only the slow pion provides sufficient kinematic constraints to reconstruct this decay.  While this technique provides an efiiciency approximately 5 times higher, it does suffer from
far higher backgrounds.

\begin{figure}[tb]
\begin{center}
\epsfig{file=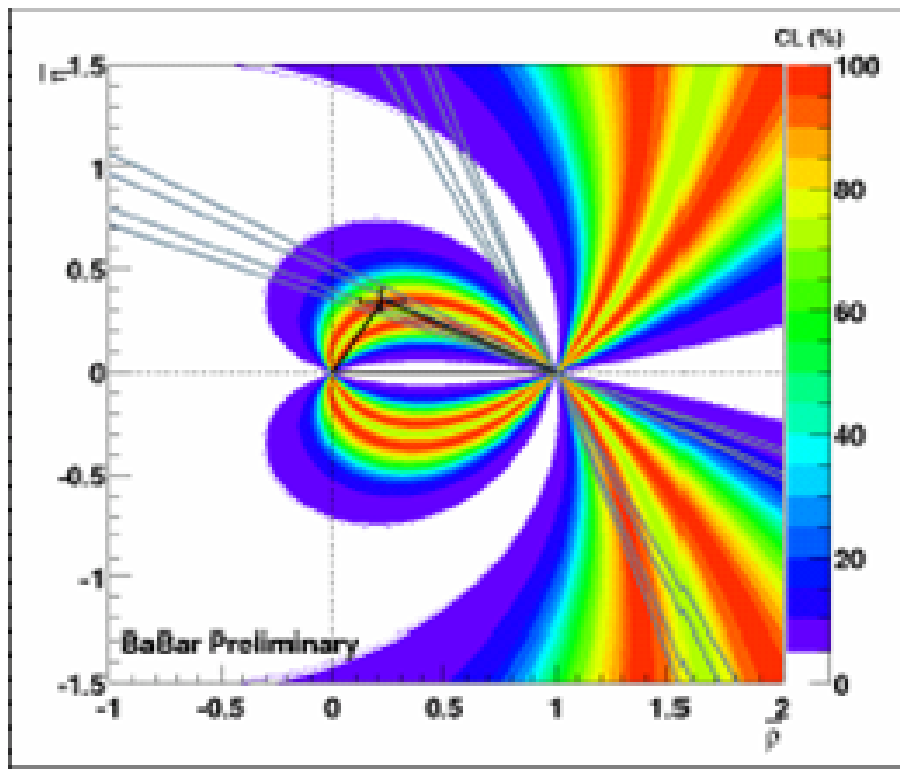,height=4.0cm}
\hspace*{1cm}
\epsfig{file=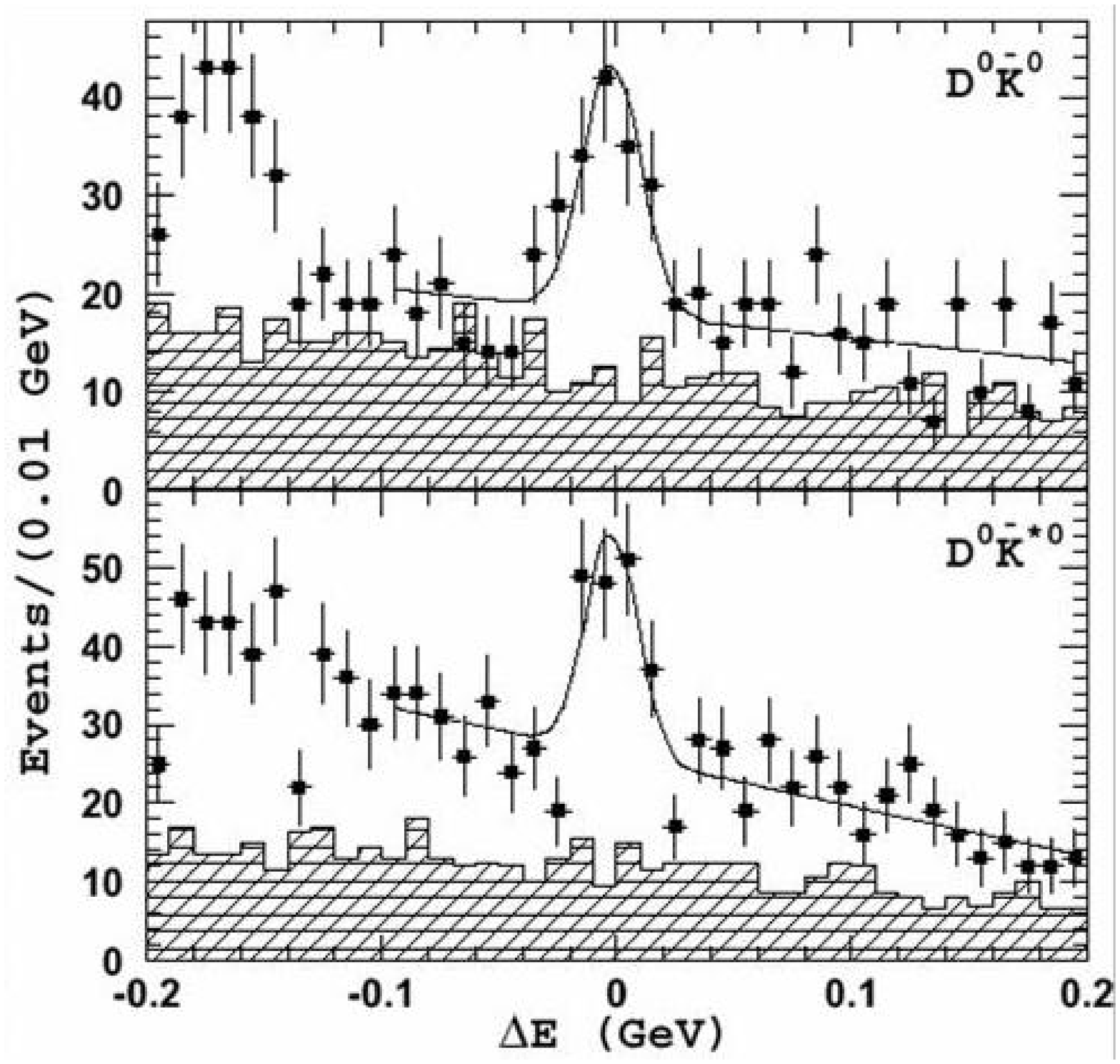,height=4.0cm}
\caption{(Left) Constraints on the $\bar{\rho}-\bar{\eta}$ plane from \babar\ \Bz $\to D^{(*)}\pi$ 
partial reconstruction results~\cite{sin2bgdpiBabarpartial}. 
(Right) Belle yields for the decay modes \Bzb $\to D^{0}\bar{K}^{0}$ and
\Bzb $\to D^{0}\bar{K}^{*0}$, where they obtain branching fractions of $(3.72 \pm 0.65 \pm 0.37) \times 10^{-5}$
and $(3.08 \pm 0.56 \pm 0.31) \times 10^{-5}$ respectively~\cite{dzkzBelle}.
}
\label{fig:timedepgam}
\end{center}
\end{figure}

The experimental observables are the coefficients of the $\sin$ and $\cos(\Delta M t)$ asymmetry terms in the time-dependent asymmetries of  \Bz(\Bzb) $\to D^{(*)}\pi$ and $D^{(*)}\rho$.
The coefficient for the $\sin$ term is equal to $2r\sin(2\beta + \gamma)\cos\delta$, where $r$ is the ratio of Cabibbo-favord and doubly-Cabibbo-suppressed amplitudes to the 
final state, and $\delta$ is the strong phase between those amplitudes.  The coefficient for the $\cos$ term is equal to $2r\cos(2\beta + \gamma)\sin\delta$.  \babar\ obtains the results:
\begin{eqnarray}
2r\sin(2\beta + \gamma)\cos\delta & = & -0.032 \pm 0.031 \pm 0.020\nonumber\\
2r\cos(2\beta + \gamma)\sin\delta & = & -0.059 \pm 0.033 \pm 0.033 \nonumber\\
2r_{*}\sin(2\beta + \gamma)\cos\delta_{*} & = & -0.049 \pm 0.031 \pm 0.020\nonumber\\
2r_{*}\cos(2\beta + \gamma)\sin\delta_{*} & = & 0.044 \pm 0.054 \pm 0.033 \nonumber
\end{eqnarray}
using exclusive reconstruction on a sample of $110 \times 10^{6}$ \BB events~\cite{sin2bgdpiBabarexcl}, 
where the first two values are from the $D\pi$ channel and the last two are from $D^{*}\pi$, and
\begin{eqnarray}
2r_{*}\sin(2\beta + \gamma)\cos\delta_{*} & = & -0.041 \pm 0.016 \pm 0.010\nonumber\\
2r_{*}\cos(2\beta + \gamma)\sin\delta_{*} & = & -0.015 \pm 0.036 \pm 0.019 \nonumber
\end{eqnarray}
using partial reconstruction on a sample of $178 \times 10^{6}$ \BB events~\cite{sin2bgdpiBabarpartial}.  Belle obtains
\begin{eqnarray}
2r\sin(2\beta + \gamma)\cos\delta & = & -0.062 \pm 0.037 \pm 0.018\nonumber\\
2r\cos(2\beta + \gamma)\sin\delta & = & -0.025 \pm 0.037 \pm 0.018 \nonumber\\
2r_{*}\sin(2\beta + \gamma)\cos\delta_{*} & = & 0.060 \pm 0.040 \pm 0.017\nonumber\\
2r_{*}\cos(2\beta + \gamma)\sin\delta_{*} & = & 0.049 \pm 0.040 \pm 0.019 \nonumber
\end{eqnarray}
using exclusive reconstruction~\cite{sin2bgdpiBelleexcl} and
\begin{eqnarray}
2r_{*}\sin(2\beta + \gamma)\cos\delta_{*} & = & -0.031 \pm 0.028 \pm 0.018\nonumber\\
2r_{*}\cos(2\beta + \gamma)\sin\delta_{*} & = & -0.004 \pm 0.028 \pm 0.018 \nonumber
\end{eqnarray}
using partial reconstruction~\cite{sin2bgdpiBellepartial}, both on a sample of $152 \times 10^{6}$ \BB events.  \babar\ obtains 
constraints on the 
value of $|\sin(2\beta + \gamma)|$ from the partial reconstruction
method: $|\sin(2\beta + \gamma)| > 0.75$ at 68\% c.l.~and $> 0.58$ at 90\% c.l., 
resulting in constraints on the $\bar{\rho}-\bar{\eta}$ plane as shown in Fig.~\ref{fig:timedepgam} left.
Belle does not claim any constraints on the value of $|\sin(2\beta + \gamma)|$ however, assuming the strong phase $\delta = 0$ or $\pi$, Belle obtains 
$2r_{*}\sin(2\beta + \gamma) = 0.031 \pm 0.028 \pm 0.013$.

The relative weak phase of the \Bz $\to D^{(*)0}K^{*0}$ and \Bzb $\to D^{(*)0}\bar{K}^{*0}$ is also
$\gamma$.  As these final states are non \CP-eigenstates, there is also a strong phase between the two decays, that can be solved for by 
additionally measuring the
\CP asymmetry in the decays \Bz $\to \bar{D}^{(*)0}K^{*0}$ and \Bzb $\to \bar{D}^{(*)0}\bar{K}^{*0}$~\cite{dzkz}.
The sensitivity to $\sin(2\beta + \gamma)$ from these decays is given by the value of 
$r \equiv \frac{|A(\Bzb \to \bar{D}^{(*)0}\bar{K}^{*0})|}{|A(\Bzb \to D^{(*)0}\bar{K}^{*0})|}$.  \babar\ obtains the following 
branching fractions~\cite{dzkzBabar}:
\begin{eqnarray}
\mathcal{B}(\Bzb \to D^{0}\bar{K}^{0}) & = & (6.2 \pm 1.2 \pm 0.4) \times 10^{-5}\nonumber\\
\mathcal{B}(\Bzb \to D^{*0}\bar{K}^{0}) & = & (4.5 \pm 1.9 \pm 0.5) \times 10^{-5}\nonumber\\
\mathcal{B}(\Bzb \to D^{0}\bar{K}^{*0}) & = & (6.2 \pm 1.4 \pm 0.6) \times 10^{-5}\nonumber
\end{eqnarray}
but obtains just a limit on the numerator of $r$:
\begin{eqnarray}
\mathcal{B}(\Bzb \to \bar{D}^{0}\bar{K}^{*0}) & < & 4.1 \times 10^{-5}\;\mbox{at 90\% c.l.} \nonumber
\end{eqnarray}
Similarly, Belle obtains~\cite{dzkzBelle}:
\begin{eqnarray}
\mathcal{B}(\Bzb \to D^{0}\bar{K}^{0}) & = & (3.72 \pm 0.65 \pm 0.37) \times 10^{-5}\nonumber\\
\mathcal{B}(\Bzb \to D^{*0}\bar{K}^{0}) & = & (3.18^{+1.25}_{-1.12} \pm 0.32) \times 10^{-5}\nonumber\\
\mathcal{B}(\Bzb \to D^{0}\bar{K}^{*0}) & = & (3.08 \pm 0.56 \pm 0.31) \times 10^{-5}\nonumber\\
\mathcal{B}(\Bzb \to D^{*0}\bar{K}^{*0}) & < & 4.8 \times 10^{-5};\mbox{at 90\% c.l.}\nonumber\\
\mathcal{B}(\Bzb \to \bar{D}^{0}\bar{K}^{*0}) & < & 0.4 \times 10^{-5}\;\mbox{at 90\% c.l.} \nonumber\\
\mathcal{B}(\Bzb \to \bar{D}^{*0}\bar{K}^{*0}) & < & 1.9 \times 10^{-5}\;\mbox{at 90\% c.l.} \nonumber
\end{eqnarray}
No constraints on $\sin(2\beta + \gamma)$ can be obtained so far from \Bz $\to D^{(*)0}(\bar{D}^{(*)0})K^{*0}$.

\section{\boldmath Using $B \to D_{(s)}^{(*)}D^{(*)}$ Decays to Measure $\gamma$}

One can combine information from $B \to D^{(*)} \Dbar^{(*)}$ and $B \to D^{(*)}_S \Dbar^{(*)}$ branching
fractions, along with \CP asymmetry measurements in $B \to D^{(*)} \Dbar^{(*)}$, to obtain a measurement of the Unitarity Triangle
angle $\gamma$~\cite{ADL,DL}.
The weak phase $\gamma$ 
may be extracted by using an SU(3) relation between the $B \to D^{(*)} \Dbar^{(*)}$ and $B \to D^{(*)}_S \Dbar^{(*)}$ decays.
In this technique, the breaking of SU(3) is parametrized via the ratios of decay
constants $f_{D_s^{(*)}}/f_{D^{(*)}}$, which are quantities measured in lattice QCD~\cite{lattice}.

One obtains the relation (for $\Bz \to \Dp\Dm$ and individual helicity states of $\Bz \to \Dstarp\Dstarm$):
\begin{equation}
\mathcal{A}_{ct}^2 = \frac{a_R \cos(2\beta + 2\gamma) - a_{\rm indir}\sin(2\beta + 2\gamma) - B}{\cos 2\gamma - 1}
\label{maineq}
\end{equation}
where
\begin{eqnarray}
B & \equiv & \frac{1}{2}(|A^D|^2 + |\bar{A}^D|^2) = \mathcal{A}_{ct}^2 + \mathcal{A}_{ut}^2 +
             2\mathcal{A}_{ct}\mathcal{A}_{ut}\cos\delta\cos\gamma , \nonumber\\
a_{\rm dir} & \equiv & \frac{1}{2}(|A^D|^2 - |\bar{A}^D|^2) = -2\mathcal{A}_{ct}\mathcal{A}_{ut}\sin\delta\sin\gamma ,\\
a_{\rm indir} & \equiv & \Im(e^{-2i\beta}A^{D*}\bar{A}^D) = -\mathcal{A}_{ct}^{2}\sin 2\beta - \nonumber\\
        & &     2\mathcal{A}_{ct}\mathcal{A}_{ut}\cos\delta\sin(2\beta + \gamma) -  \mathcal{A}_{ut}^2\sin(2\beta + 2\gamma) ,\nonumber
\end{eqnarray}
and
\begin{equation}
a_R^2 \equiv B^2 - a_{\rm dir}^2 - a_{\rm indir}^2.
\end{equation}
$B$ represents the branching fraction to a given $B \to D^{(*)} \Dbar^{(*)}$ decay and $a_{\rm dir}$ and $a_{\rm indir}$
represent the corresponding direct and indirect \CP asymmetries respectively.
The phases $\beta$ and $\gamma$ are the angles of the Unitarity Triangle and $\delta$ is a strong phase.
$\mathcal{A}_{ct} \equiv |(T + E + P_c - P_t - P_{EW}^C)V_{cb}^*V_{cd}|$ and
$\mathcal{A}_{ut} \equiv |(P_u - P_t -P_{EW}^C)V_{ub}^*V_{ud}|$ are the norms of the combined $B \to D^{(*)} \Dbar^{(*)}$ decay amplitudes containing
$V_{cb}^*V_{cd}$ and $V_{ub}^*V_{ud}$ terms respectively, and the $T$, $P$, and $E$ terms are the
tree, penguin, and exchange amplitude components respectively~\cite{DL}.  As the modes $\Bz \to \Dstarpm\Dmp$ are not \CP eigenstates,
a slightly more complicated formalism is needed for these modes~\cite{ADL}.

Using these relations, there are 5 variables for each $B \to D^{(*)} \Dbar^{(*)}$ decay for which to solve: $\mathcal{A}_{ct}$,
$\mathcal{A}_{ut}$, $\delta$, $\beta$, and $\gamma$.  The branching fraction and the direct and indirect
\CP asymmetries of the $B \to D^{(*)} \Dbar^{(*)}$ decay provide three measured quantities.  Two further quantities are needed.
The angle $\beta$ can be obtained from the measurements of $\sin 2\beta$ using charmonium ``golden modes''
at the $B$-factories~\cite{HFAG}.  The other measurement that can be used is the branching fraction of the corresponding
$B \to D^{(*)}_S \Dbar^{(*)}$ decay.

If SU(3) were a perfect symmetry between $B \to D^{(*)} \Dbar^{(*)}$ and $B \to D^{(*)}_S \Dbar^{(*)}$ decays, then the norm of the $B \to D^{(*)}_S \Dbar^{(*)}$
amplitude, denoted $\mathcal{A}_{ct}^{\prime}$, would equal $\mathcal{A}_{ct}/\sin\theta_c$, where $\theta_c$
is the Cabibbo angle.  However, SU(3)-breaking effects can spoil this relation.  
The SU(3)-breaking can be parametrized by the ratio of decay constants $f_{D_s^{(*)}}/f_{D^{(*)}}$, such that
$\mathcal{A}_{ct}^{\prime} = f_{D_s^{(*)}}/f_{D^{(*)}} \times \mathcal{A}_{ct}/\sin\theta_c$ (where the parentheses
around the asterisks correspond to the $B \to D^{(*)} \Dbar^{(*)}$ and $B \to D^{(*)}_S \Dbar^{(*)}$ decays that are used).     
The theoretical uncertainty of this relation is determined to be 10\%~\cite{DL}.
\begin{figure}[tb]
\begin{center}
\epsfig{file=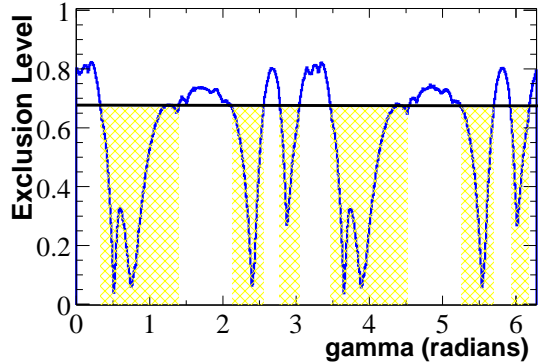,height=5.0cm}
\caption{
Constraints on the value of $\gamma$ from $B \to D_{(s)}^{(*)}D^{(*)}$ decays.
}
\label{fig:dsstdstgam}
\end{center}
\end{figure}

Using these relations, together with measurements of $B \to D^{(*)} \Dbar^{(*)}$ and $B \to D^{(*)}_S \Dbar^{(*)}$ branching fractions and
\CP asymmetries from \babar\ and Belle, constrains $\gamma$ to lie in one of the ranges $[19.4^{\circ}-80.6^{\circ}]$, $[120^{\circ}-147^{\circ}]$,
or $[160^{\circ}-174^{\circ}]$ at 68\% confidence level.  There is an additional $(+0^{\circ} \mbox{ or } 180^{\circ})$ phase ambiguity for each range.
The constraints
disappear for larger confidence levels, however the \babar\ measurements used for these constraints were obtained on on a sample of only 
$88 \times 10^{6}$ \BB events and thus can be significantly improved.

\section{Conclusions}

Although the angle $\gamma$ is the most difficult to measure of the Unitarity Triangle angles at the $B$-Factories, surprising
progress has been made in constraining it over the past few years.  We now have initial measurements of the values of both $\gamma$ and 
$\sin(2\beta + \gamma)$ from multiple channels, and have progressed toward precision measurements of this angle, which appear poised to 
have errors below $\pm10^{\circ}$ prior to physics at the LHC.  These precision measurements of $\gamma$ are a critical test
for the consistency of the Standard Model mixing sector.

\section*{Acknowledgments}
As a member of the \babar\ Collaboration, I thank my \babar\ and PEP-II colleagues
for their support and their tremendous dedicated efforts.  I congratulate the FPCP'04 committees
for organizing an excellent conference.
The author was partially supported by DOE contract DE-FG01-04ER04-02 and by a Caltech Millikan 
Fellowship.

%
\label{JAlbertEnd}

\end{document}